\documentclass[11pt]{article}
\usepackage{amssymb}
\usepackage{amsmath}
\usepackage{accents}
\usepackage{texdraw}

\newcommand{\dif}{\mbox{\rm{d}}}
\newcommand{\ii}{\mbox{\rm{i}}}
\newcommand{\wh}{\widehat}
\newcommand{\wt}{\widetilde}

\DeclareMathAccent{\wtilde}{\mathord}{largesymbols}{"65}
\DeclareMathAccent{\what}{\mathord}{largesymbols}{"62}

\makeatletter
\def\wb{\accentset{{\cc@style\underline{\mskip10mu}}}}
\makeatother

\begin{document}

\title{\bf The Boussinesq integrable system.\\
Compatible lattice and continuum structures}
\author{Anastasios Tongas and Frank Nijhoff \\
Department of Applied Mathematics, University of Leeds, Leeds LS2 9JT, U.K.}

\maketitle

\begin{abstract}

We consider the discrete Boussinesq integrable system and the compatible set of
differential difference, and partial differential equations. The latter
not only encode the complete hierarchy of the Boussisesq equation, but also 
incorporate the hyperbolic Ernst equations for an Einstein-Maxwell-Weyl field in
general relativity. We demonstrate a specific symmetry reduction of the partial 
differential equations, to a six-parameter, second order coupled system of ordinary 
differential equations, which is conjectured to be of Garnier type. 
\end{abstract}

\section{Introduction}
  
One of the outstanding universal features of integrable nonlinear evolution 
equations is their appearance together with a structured infinite hierarchy of 
commuting flows, i.e. an infinite set of {\em mutually} compatible partial 
differential equations (PDE). This feature reflects the extremely high degree 
of symmetry hidden in the system, and it has been exploited in many ways 
leading to the intimate connection between the associated hierarchies on the 
one hand, and the representation theory of infinite-dimensional Lie algebras 
and loop groups, on the other hand, cf \cite{Kac}, \cite{miwa} and references
therein. However, the point of view that one single equation in the associated 
hierarchy possesses a dominant role is misguided and, in fact, one should 
consider the infinite set of PDEs altogether, which constitutes the hierarchy, 
as being the integrable system.    
   
It is remarkable that there are situations in which the complete infinite 
hierarchy of flows of an integrable evolution equation, can be encoded in a 
single PDE. A prime example of such an equation was given recently in 
\cite{NHJ}, and it is represented be a fourth order scalar PDE. The complete 
hierarchy of the Korteweg--de Vries (KdV) equation follows by systematic 
expansions from this PDE. In the present work we consider  another 
{\em hierarchy generating} PDE, namely the one that encodes the complete 
hierarchy of the Boussinesq (BSQ) equation. This will be achieved by deriving, 
first, the compatible discrete integrable system.  	 

The construction of the latter is based on the infinite-matrix structures, 
introduced in \cite{Nij1}. In particular we consider those structures which 
serve as a representation of the loop algebra built from the hierarchy of flows,
of the second member in the Gel'fand-Dikii hierarchy, namely the Boussinesq 
equation \cite{NijPap}. The basic ingredient consists of a linear integral
equation for (infinite) basic functions depending on a complex parameter and on 
the coordinates of the system. The coordinates can be chosen to be discrete as 
well as continuous, and exactly this freedom in the coordinate choice, is 
responsible for making the integral equations such a convenient tool to derive 
discrete integrable systems. This is the subject matter of section 2, which ends
up with a parameter family of discrete equations, namely the lattice BSQ 
equation.

In section 3, the multi-dimensional consistency of the lattice BSQ equation is 
tested, by embedding the two dimensional lattice into a higher dimensional 
lattice. This aspect of the discussion forms a completion to the study of the 
lattice BSQ integrability as performed in \cite{Wphd}. The fact that the 
evolution in the extended three dimensional lattice is well posed reveals, in 
an essentially algorithmic fashion, the Lax pair of the lattice BSQ equation. 
Section 4, proceeds with the derivation of the compatible set of 
differential-difference equations of the lattice BSQ equation. This will be 
achieved by imposing on the dependent variables an additional dependence on the 
continuous lattice parameters, in such a way that certain compatibility 
conditions are satisfied.
    
Sections 5 and 6, deal with the continuous compatible PDEs and ODEs, 
respectively. First, we present the generating PDE of the Boussinesq hierarchy, 
consisting of a two-parameter family of coupled fourth order system of PDEs, in 
two dependent and two independent variables. Next, we illustrate the connection 
of the latter PDEs with the Boussinesq hierarchy, using the Lax pair of the 
relevant PDEs, which was derived in the preceding sections. A prominent 
position, among the subsystems which are included into the solution space of 
fourth order parameter family of PDEs, is possessed by a second order system.  
It is the heart of the Einstein's field equations for plane symmetric 
spacetimes, in the presence of certain massless fields. In the context of 
general relativity, the latter equations are known as the Ernst equations for an
Einstein-Maxwell-Weyl field. Finally, we consider a particular type of 
similarity solutions of the system of PDEs under consideration, which are built 
from the solutions of a second order coupled system of ODEs, involving six free 
parameters. The latter may be considered to be associated with the hierarchies 
of the Painlev\'e VI equations, 

The paper concludes with perspectives, where various avenues for future research
are discussed.
   
\section{The discrete Boussinesq system}

In order to derive the lattice Boussinesq (BSQ) system, we shall use a linear
integral equation of the following form
\begin{equation}
{\boldsymbol u}(k) + \int_{C} \dif \lambda(\ell) 
\frac{{\boldsymbol u}(k)}{k-\omega \ell} = 
\rho(k) {\boldsymbol r}(k)\,. \label{eq:int}
\end{equation}
cf \cite{NijPap}. The vector ${\boldsymbol u}(k)$, has $i \in \mathbb Z$ 
entries, each of them taking values in the complex numbers, and 
${\boldsymbol r}$ is a vector with entries given by ${r}_i=k^i$, 
$(i \in \mathbb{Z})$. The variables $\ell$ and $k$ are complex, and $\omega$ is 
a $N$th root of unity, i.e. $\omega=\exp (2 \ii \pi/N)$. The integrations in 
(\ref{eq:int}) are performed over an arbitrary contour $C$ in the complex 
$k$-plane, and over an arbitrary measure $\dif \lambda (\ell)$. In principle, 
the only restrictions on the measure and the contour consists of the requirement
of the existence and uniqueness of the solution of the integral equation 
(\ref{eq:int}), once the inhomogeneous term is specified. Along with the 
infinite vector ${\boldsymbol u}$, which solves the integral equation
(\ref{eq:int}), we can introduce an infinite-component matrix potential 
$\boldsymbol{U}$, given by
\begin{equation}
{\boldsymbol U}=\int_{C} \dif \lambda(\ell)\, {\boldsymbol u}(\ell)\,\,  
\phantom{}^t{\boldsymbol r(\ell)} \,,
\label{eq:defU}
\end{equation}      
where $\phantom{}^t{\boldsymbol r}$ denotes the transposed vector of 
${\boldsymbol r}$. The aim is to derive for the matrix potential 
$\boldsymbol{U}$, as well as for the vector potential $\boldsymbol{u}$, a set 
of (infinite-) matrix equations, by imposing on these objects a dependence on 
additional variables and certain transformation rules. The variables can be 
continuous as well as discrete. Adopting the latter case, we consider two 
operations, namely the maps 
\begin{equation}
{\boldsymbol u}(n,m,k) \mapsto \widetilde{\boldsymbol u}(n,m,k)= 
{\boldsymbol u}(n+1,m,k),\qquad {\boldsymbol u}(n,m,k) 
\mapsto \widehat{\boldsymbol u}(n,m,k)={\boldsymbol u}(n,m+1,k)\,. 
\label{eq:maps}
\end{equation}
These maps have the interpretation of shifts in a two-dimensional lattice, 
but equally well of B\"acklund transformations generated by a 
transformation of the complex function $\rho(k)$, which is not specified yet.
The transformation rules which are imposed on $\rho(k)$ are
\begin{equation}
\rho(k;p,q) \mapsto \widetilde{\rho}(k;p,q) = 
\frac{p+k}{p+\omega k}\,\rho(k;p,q)\,,\quad
\rho(k;p,q) \mapsto \widehat{\rho}(k;p,q) = 
\frac{q+k}{q+\omega k}\,\rho(k;p,q)\,, \label{eq:BT}
\end{equation}
where $p$ and $q$ are the complex lattice parameters, associated with the 
lattice variables $n$ and $m$, respectively. Using the linear integral equation 
(\ref{eq:int}), we may determine how $\boldsymbol{u}$ changes, under the action 
of the transformations (\ref{eq:BT}). Restricting to the special case $N=3$, 
i.e. $\omega=\exp (2 \ii \pi/3)$, the equations which determine the change of 
$\boldsymbol{u}$, are given by 
\begin{subequations}\label{eq:infu}
\begin{eqnarray}
(p+\omega k) {\widetilde{\boldsymbol u}}&=& (p+{\boldsymbol \Lambda} - 
{\widetilde {\boldsymbol U}}\, {\boldsymbol O})\,
{\boldsymbol u}\,, \label{eq:infu1}\\
(p+k)(p+\omega^2 k) {\boldsymbol u}
&=&(p+\omega {\boldsymbol \Lambda})(p+\omega^2 {\boldsymbol \Lambda})\, 
\widetilde{\boldsymbol u} \nonumber \\ 
& &-{\boldsymbol U}\,\left(\omega {\boldsymbol O} \, 
(p+\omega^2 {\boldsymbol \Lambda})+\omega^2 (p+\omega^2\,\, 
\phantom{}^t{\!\boldsymbol \Lambda})\,{\boldsymbol O}\right)\, 
{\widetilde{\boldsymbol u}}\,, \label{eq:infu2} 
\end{eqnarray}
\end{subequations}
together with the same equations by replacing 
$\phantom{}\,\,\widetilde{\cdot}\,\,$ and $p$, with 
$\phantom{}\,\,\widehat{\cdot}\,\,$ and $q$, respectively. The matrices 
${\boldsymbol \Lambda}$ and $\phantom{}^t{\boldsymbol \Lambda}$, define  the 
operations of index-raising when multiplied from the left and from the right, 
respectively, i.e. (${\boldsymbol \Lambda}\, {\boldsymbol u})_i=u_{i+1}$. The 
matrix ${\boldsymbol O}$ is a projection matrix which singles out the central 
element of the object which upon acts, i.e. 
$({\boldsymbol O}\, {\boldsymbol u})_i=u_0 \delta_{i0}$, 
where $\delta_{ij}$ the Kronecker delta symbol. Integrating (\ref{eq:infu}) over
the contour $C$ and the same measure $\dif \lambda(\ell)$, we obtain the 
following nonlinear system of equations for the matrix potential 
${\boldsymbol U}$ 
\begin{subequations} \label{eq:infU1}
\begin{eqnarray}
{\widetilde{\boldsymbol U}}(p+\omega \,\,\phantom{}^t{\!\boldsymbol \Lambda}) 
&=& (p+{\boldsymbol \Lambda} - 
\widetilde{\boldsymbol U}\, {\boldsymbol O})\, {\boldsymbol U}\,, 
\label{eq:infU11} \\
{\boldsymbol U}\, 
(p+\phantom{}^t{\!\boldsymbol \Lambda})
(p+\omega^2\,\,\phantom{}^t{\!\boldsymbol \Lambda}) &=&
(p+\omega {\boldsymbol \Lambda})(p+\omega^2\,\, {\!\boldsymbol \Lambda})\, 
{\widetilde {\boldsymbol U}} \nonumber \\ 
& &-{\boldsymbol U}\,\left(\omega {\boldsymbol O} \, 
(p+\omega^2 {\boldsymbol \Lambda})+\omega^2 (p+\omega^2\,\, 
\phantom{}^t{\!\boldsymbol \Lambda})\,{\boldsymbol O}\right)\, 
{\widetilde{\boldsymbol U}}\,. \label{eq:infU12} 
\end{eqnarray}
\end{subequations}
Using the identity $\omega^2+\omega+1=0$, the above equations may be rewritten 
as follows
\begin{subequations} \label{eq:infU2}
\begin{eqnarray}
{\widetilde{\boldsymbol U}}(p+\omega \,\,\phantom{}^t{\!\boldsymbol \Lambda}) 
&=& (p+{\boldsymbol \Lambda} - \widetilde{\boldsymbol U}\, {\boldsymbol O})\, 
{\boldsymbol U} \label{eqU1}\,, \label{eq:infU21} \\
{\boldsymbol U}\, \left(p^2 - \omega p\,\,\phantom{}^t{\!{\boldsymbol \Lambda}}+ 
\omega^2\,(\phantom{}^t{\!{\boldsymbol \Lambda}})^2\right) &=& 
(p^2 - p\, {\boldsymbol \Lambda} + {\boldsymbol \Lambda}^2) \, 
{{\boldsymbol{\wt U}}} + {\boldsymbol U}\,\left( p\,{\boldsymbol O} - 
{\boldsymbol O}\, {\boldsymbol \Lambda} - \omega\,\,\phantom{}^t{\!\boldsymbol 
\Lambda }\, {\boldsymbol O}\right)\,
{\widetilde {\boldsymbol U}}\,. \qquad \label{eq:infU22}
\end{eqnarray}
\end{subequations}
From the above infinite set of relations for the components of the potential 
${\boldsymbol U}$, we restrict ourselves to the equations evolving its $(0,0)$, 
$(1,0)$ components from eq. (\ref{eq:infU21}) and the $(0,0)$ component from 
eq. (\ref{eq:infU22}), together with the same equations by interchanging 
$\phantom{}\,\,\widetilde{\cdot}\,\,$ and $p$, with 
$\phantom{}\,\,\widehat{\cdot}\,\,$ and $q$, respectively. A simple elimination 
process, using the discrete operations, leads to a closed system of equations 
for the variables $(U_{00},U_{10},U_{01})$. The omitted latter set of discrete 
equations, obtains a simple form by introducing new dependent variables 
$(u,v,w)$ by the relations
\begin{subequations} \label{eq:newvar}
\begin{eqnarray}
u&=&\textstyle{\phantom{-\omega}U_{00} - (n\, p+m\, q)}\,, \\
v&=&\textstyle{\phantom{-\omega}U_{10} - (n\, p+m\, q) U_{00} + 
\frac{1}{2}n(n+1)p^2 + n\, m\, p\, q + \frac{1}{2}m(m+1)q^2}\,, \\
w&=&\textstyle{-\omega U_{01} - (n\, p+ m\, q) U_{00} + 
\frac{1}{2}n(n-1)p^2 + n\, m\, p\, q + \frac{1}{2}m(m-1)q^2} \,.
\end{eqnarray}
\end{subequations}         
In terms of these new variables, the aforementioned set of partial difference 
equations (P$\Delta$E), reads
\begin{subequations}\label{eq:latticeBSQ}
\begin{eqnarray} 
{\widetilde w} &=& u {\widetilde u} - v\,, \label{eq:latticeBSQwt} \\  
{\widehat w} &=& u {\widehat u} - v\,, \label{eq:latticeBSQwh} \\
w &=& u {\wh {\wt u}} - {\wh {\wt v}} + \frac{\tau-\sigma}{\wh u - \wt u}\,, 
\label{eq:latticeBSQvth} 
\end{eqnarray}
\end{subequations}
where $\tau=p^3$, $\sigma=q^3$. The system of P$\Delta$Es (\ref{eq:latticeBSQ}), 
will be referred in the following as the lattice BSQ system.

\section{Three dimensional consistency of the lattice Boussinesq}

The lattice BSQ system (\ref{eq:latticeBSQ}), represents a {\em compatible 
parameter family} of P$\Delta$Es, in the sense that it can be embedded 
consistently in a multidimensional lattice, on which the evolution is well 
posed, \cite{NW}. Recently, this property has been successfully exploited 
for quadrilateral {\em scalar} lattice equations to obtain a classification
of the latter, \cite{Adler}.     

In order to illustrate the above statement, we extend the lattice into a third 
dimension by introducing a new lattice variable $l$ associated with a new shift,
which will be denoted by $\phantom{.}{\wb \cdot}\,$, and a new complex parameter
$\zeta$. Embedding the system of equations (\ref{eq:latticeBSQ}) into the 
generated three dimensional lattice, we get      
\begin{subequations} \label{eq:3DBSQ}
\begin{eqnarray}
{\wt w} &=& u {\wt u} - v,\qquad w = u {\wh {\wt u}} - 
{\wh {\wt v}} + \frac{\tau-\sigma}{\wh u - \wt u} \,, 
\label{eq:3Dwtw}\\  
{\wh w} &=& u {\wh u} - v,\qquad w = u {\wb {\wt u}} - {\wb {\wt v}} + 
\frac{\tau-\zeta}{{\wb u} - {\wt u}} \,, \label{eq:3Dwhw}\\
{\wb w} &=& u {\wb u} - v,\qquad w = u {\wb {\wh u}} - {\wb {\wh v}} + 
\frac{\sigma-\zeta}{{\wb u} - {\wh u}}\,.
\label{eq:3Dwbw} 
\end{eqnarray}
\end{subequations}
Using from (\ref{eq:3DBSQ}), those equations which involve the shift of the 
variable $w$ in one direction, and taking into account the relevant 
compatibility conditions, we can determine how $u$ and $w$, are shifted in two 
directions. The corresponding equations for the variable $u$ read 
\begin{equation} \label{eq:u2shifts}
{\wh {\wt u}} = \frac{{\wt v}-{\wh v}}{{\wt u}-{\wh u}}\,, \qquad 
{\wb {\wt u}} = \frac{{\wb v}-{\wt v}}{{\wb u}-{\wt u}}\,, \qquad
{\wb {\wh u}} = \frac{{\wb v}-{\wh v}}{{\wb u}-{\wh u}}\,. 
\end{equation}

\begin{figure}[h]
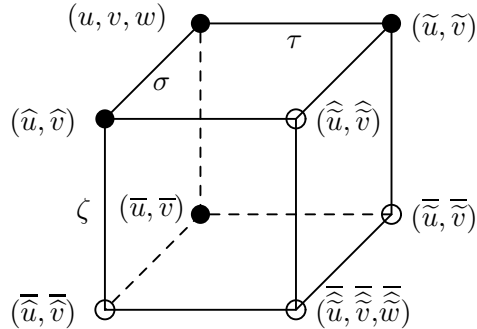

\centertexdraw{ \setunitscale 1. \linewd 0.01
\move (0 0) \linewd 0.01 \lpatt(0.05 0.05)
\lvec(0.0 1.0) \lpatt() \lvec (1.0 1.0) \linewd 0.01 \lvec (1.0 0.0)  
\lpatt(0.05 0.05) \lvec(0 0)  
\lvec(-0.5 -0.5) \lpatt()  \lvec(-0.5 0.5)  \linewd 0.01 \lvec(0 1.0) 
\move (1. 0)  \linewd 0.01
\lvec(0.5 -0.5) \lvec(-0.5 -0.5) \move(1.0 1.0)  
\lvec(0.5 0.5) \lvec(-0.5 0.5) \move(0.5 0.5) \lvec(0.5 -0.5)
\move(0 0) \fcir f:0.0 r:0.05 \move(1 1) \fcir f:0.0 r:0.05 \move(-0.5 0.5) 
\fcir f:0.0 r:0.05
\move(0 1) \fcir f:0.0 r:0.05
\move(-0.5 -0.5) \lcir r:0.05 \move(0.5 0.5) \lcir r:0.05 
\move(0.5 -0.5) \lcir r:0.05
\move(1 0) \lcir r:0.05  
\htext (-0.7 0.95) {$(u,v,w)$} 
\htext (-0.43 -0.05) {$(\wb{u},\wb{v})$} 
\htext (1.1 0.9) {$({\wt{u}},{\wt v})$}
\htext (-1. 0.4) {$({\wh u},{\wh v})$} 
\htext (0.6 -0.6) {$({\wb {\wh {\wt u}}},{\wb {\wh {\wt v}}},{\wb {\wh {\wt{\!w}}}})$}
\htext (1.1 -0.1) {$({\wb {\wt u}},{\wb {\wt v}})$}
\htext (-1. -0.6) {$({\wb {\wh u}},{\wb {\wh v}})$}
\htext (0.6 0.4) {$({\wh {\wt u}},{\wh {\wt v}})$}
\htext (-0.25 0.65) {$\sigma$} \htext (0.45 0.87) {$\tau$} 
\htext (-0.65 -0.05) {$\zeta$}
}
\caption{Three dimensional consistency of the lattice Boussinesq system.}
\label{fig:3D}
\end{figure}
  
Let us now impose the following initial values
\begin{equation}
(u,v,w),\quad ({\wt u}, {\wt v}), \quad ({\wh u}, {\wh v}), \quad ({\wb u}, {\wb v}),  
\end{equation}
at the vertices depicted by $\bullet$ in Figure \ref{fig:3D}. Successively, 
using equations (\ref{eq:3DBSQ}), (\ref{eq:u2shifts}) we may evaluate the 
updates
\begin{equation}
({\wh {\wt u}}, {\wh {\wt v}}),\quad ({\wb {\wt u}}, {\wb {\wt v}}), 
\quad ({\wb {\wh u}}, {\wb {\wh v}})\,,  
\end{equation}
which are assigned to the vertices depicted with $\circ$. Then, from these data 
and the evolution of the lattice equations (\ref{eq:3DBSQ}), (\ref{eq:u2shifts}) 
we are free to calculate the values at the final point, in three different ways.
In principle, there is no particular reason for these three values to coincide, 
unless the P$\Delta$E's (\ref{eq:latticeBSQ}) enjoy this special property. The 
confirmation of this test is crucial in that it verifies, at first stage, the 
consistency of the embedding of the lattice BSQ into a multidimensional lattice. 
By simple and straightforward algebraic calculations, and eliminating all terms 
shifted in two directions in favour of terms shifted in one direction, one 
finds that  
\begin{subequations} \label{eq:finalvalue}
\begin{eqnarray}
{\wb {\wh {\wt u}}} &=& u + \frac{(\zeta-\sigma)({\wt u} - {\wb u}) - 
(\zeta-\tau)({\wh u}-{\wb u})}
{({\wh v}-{\wb v})({\wt u} - {\wb u}) - ({\wt v}-{\wb v})({\wh u}-{\wb u})}\,, 
\label{eq:uthb} \\
{\wb {\wh {\wt v}}} &=& v + \frac{(\zeta-\sigma)({\wt u} - {\wb u}){\wh u} - 
(\zeta-\tau)({\wh u}-{\wb u}){\wt u}}{({\wh v}-{\wb v})({\wt u} - {\wb u})\; 
-\; ({\wt v}-{\wb v})({\wh u}-{\wb u})\phantom{u}}\,, \label{eq:vthb} \\
{\wb {\wh {\wt{\!w}}}} &=& w + \frac{(\zeta-\sigma)({\wt v} - {\wb v}) - 
(\zeta-\tau)({\wh v}-{\wb v})}
{({\wh v}-{\wb v})({\wt u} - {\wb u}) - ({\wt v}-{\wb v})({\wh u}-{\wb u})}\,, 
\label{eq:wthb}
\end{eqnarray}
\end{subequations}
independently of the way of calculating the values at this point. This fact can 
be easily verified, since the fractional terms in equations 
(\ref{eq:finalvalue}), remain invariant under any permutation of the three 
lattice shifts, along with the corresponding lattice parameters. 

Having established the three dimensional consistency of the lattice BSQ (using 
no more information than the equations themselves), it is now straightforward to
derive the linear representation for the lattice BSQ system. This is 
accomplished by appointing $({\wb u},{\wb v})$ as projective variables, and 
identifying them with the ratio of homogeneous variables 
$(\psi_0,\psi_1,\psi_2)$, as follows
\begin{equation} 
{\wb u}=\frac{\psi_1}{\psi_0},\qquad {\wb v}=\frac{\psi_2}{\psi_0}\,. 
\label{eq:project}
\end{equation}      
Substituting $({\wb u},{\wb v})$ given by equations (\ref{eq:project}), into the
relevant equations (\ref{eq:3DBSQ}), (\ref{eq:u2shifts}) and setting
\begin{equation}
{\wt \psi_0}=-{\wt u} \psi_0 + \psi_1,\qquad {\wh \psi_0}=-{\wh u} \psi_0 + 
\psi_1\,, \label{eq:prelax1}
\end{equation}      
we obtain 
\begin{subequations} \label{eq:prelax2}
\begin{eqnarray}
{\wt \psi_1}&=&-{\wt v} \psi_0 - \psi_2\,,\qquad 
{\wt \psi_2}=(\tau-\zeta + w {\wt u}- u {\wt v})\psi_0 - w \psi_1 + u \psi_2\,,
\label{eq:prelax21} \\
{\wh \psi_1}&=&-{\wh v} \psi_0 + \psi_2\,, \qquad
{\wh \psi_2}=(\sigma-\zeta + w {\wh u}- u {\wh v})\psi_0 -w \psi_1 + u \psi_2\,.
\label{eq:prelax22}
\end{eqnarray} 
\end{subequations}
The linear system (\ref{eq:prelax1}), (\ref{eq:prelax2}) can be written in a 
matrix form as follows  
\begin{equation} \label{eq:LaxBSQ}
{\boldsymbol{\wt \psi}}=L{\boldsymbol{\psi}} \equiv \left( \begin{array}{rcc}
-{\wt u}&1&0\\
-{\wt v}&0&1\\
\triangleright&-w&u \end{array} \right) \, {\boldsymbol{\psi}} , \qquad
{\boldsymbol{\wh \psi}}=M{\boldsymbol{\psi}} \equiv \left( \begin{array}{rcc}
-{\wh u}&1&0\\
-{\wh v}&0&1\\
\triangleleft&-w&u \end{array} \right) \, {\boldsymbol{\psi}} \,,
\end{equation}
where ${\boldsymbol{\psi}}=\phantom{}^t(\psi_0,\psi_1,\psi_2)$, and the entries 
of the matrices $L$ and $M$ which are denoted by $\triangleright$ and 
$\triangleleft$, are uniquely determined by the relations
\begin{equation} 
\det L=\tau-\zeta\,,\qquad \det M=\sigma-\zeta \label{eq:detLM}\,,
\end{equation}
respectively. It should be noticed that the compatibility condition, 
${\wh L}M={\wt M}L$, of the linear system (\ref{eq:LaxBSQ}) for every value of 
$\zeta$, holds on weaker equations than the lattice BSQ system 
(\ref{eq:latticeBSQ}). The validity of the previous compatibility condition on 
the solutions of the lattice BSQ, implies that the linear difference system 
(\ref{eq:LaxBSQ}), can be solved for every fixed value of $\zeta$. Thus, we may 
replace the column vector $\boldsymbol{\psi}$, by a matrix $\Psi$ with values in
${\rm GL}(3,\mathbb{C})$, and which is formed by putting together three 
independent solutions of the system (\ref{eq:LaxBSQ}). 

\section{The differential-difference compatible system}

We now impose on $\Psi$, and subsequently on the dependent variables, an 
additional dependence  on the continuous lattice parameters $(\tau,\sigma)$.
This will be done by imposing on the linear system (\ref{eq:LaxBSQ}), certain 
first order linear differential relations for $\Psi$, in such a way that the 
following compatibility conditions 
\begin{equation}
\partial_{\tau}{\wt \Psi}={\wt{\partial_{\tau}\Psi}}\,,\qquad   
\partial_{\sigma}{\wh \Psi}={\wh{\partial_{\sigma}\Psi}}\,, \label{eq:DScomp}
\end{equation}
are satisfied. Partial differentiation will be denoted by a comma, or by 
$\partial$, followed by the variable(s) with respect to which the 
differentiation has been performed. From the linear system (\ref{eq:LaxBSQ}), 
where now $\boldsymbol{\psi}$ is replaced by the matrix $\Psi$, it follows that 
\begin{equation}
\det \Psi = (\tau-\zeta)^n (\sigma-\zeta)^m \,, \label{eq:detPsi} 
\end{equation}  
up to a non significant constant which may be fixed to unit, without losing 
generality. The previous observation and the matrix identity  
\begin{equation}
{\rm{tr}}\,(\Psi_{,\tau} \Psi^{-1}) = 
\frac{(\det \Psi)_{,\tau}}{\det \Psi\,\,\,}\,, \label{eq:tracePsi}
\end{equation}
suggest that, a potential simple form of the first order linear differential 
equations for $\Psi$, is 
\begin{equation}
{\Psi}_{,\tau}=\frac{A}{\tau-\zeta}\, \Psi\,, \qquad 
{\Psi}_{,\sigma}=\frac{B}{\sigma-\zeta}\, \Psi\,. \label{eq:contLax}
\end{equation}   
Here, $A$ and $B$ take values in ${\rm gl}(3,\mathbb{C})$, they do not depend on
$\zeta$ and
\begin{equation}
{\rm tr}\, A=n\,, \qquad{\rm tr}\, B=m\,. \label{eq:traceAB}
\end{equation} 
The compatibility condition 
$\partial_{\tau}{\wt \Psi}={\wt{\partial_{\tau}\Psi}}$, for every value of 
$\zeta$, applied on the relevant equations (\ref{eq:LaxBSQ}), 
(\ref{eq:contLax}), leads to the following system
\begin{subequations} \label{eq:semi}
\begin{eqnarray}
G_{,\tau} + \varLambda + \varLambda A - {\wt A} \varLambda &=&0\,,
\label{eq:semi1} \\
G A-{\wt A}G&=&0\,, \label{eq:semi2}
\end{eqnarray}
\end{subequations}
where 
\begin{equation}
G=L-(\tau-\zeta) \varLambda\,, \label{eq:defG}
\end{equation}
and the matrix $\varLambda$ is given by 
\begin{equation}
\varLambda = \left( \begin{array}{ccc}0&0&0\\ 0&0&0 \\ 1&0&0 \end{array} 
\right)\,. \label{eq:defLam}
\end{equation} 
After a lengthy elimination process applied to the system of equations 
(\ref{eq:semi}), using the discrete operation, and taking also into account 
equation (\ref{eq:latticeBSQwt}) and the relevant trace relation from 
(\ref{eq:traceAB}), one obtains the explicit form of the matrix $A$, and 
simultaneously the compatible system of D$\Delta$Es for the dependent variables.
It turns out that the compatibility conditions (\ref{eq:semi}), amount exactly 
to the equations obtained by identifying two different dyadic forms for the 
matrix $A$. Specifically, $A$ is given by
\begin{equation}
A_{ij}= a_i\, a'_j = c_i \, c'_j\,,  \label{eq:dyadicA}
\end{equation}
where
\begin{subequations}\label{eq:compA} 
\begin{align}
{a}_i&=(1,{\wt u},{\wt v})\,,  
&{a'}_i&=(n+{\wt v} u_{,\tau} - {\wt u} w_{,\tau}\,,\,w_{,\tau}\,,\,- 
u_{,\tau})\,,\label{eq:compA1}\\
{c}_i& = (-u_{,\tau}\,,\,-v_{,\tau}\,,\,n+{\undertilde w} u_{,\tau} - 
{\undertilde u} v_{,\tau})\,,
&{c'}_i&=({\undertilde w},-{\undertilde u},1)\,, \label{eq:compA2}
\end{align} 
\end{subequations}
$i=0,1,2$, and $\phantom{.}\undertilde{\cdot}\,$ denotes backward shift in the 
lattice variable $n$. The identification of the two dyadic forms of $A$, leads 
to the following differential-difference equations (D$\Delta$E)
\begin{equation}
v_{,\tau}={\wt u}\, u_{,\tau},\qquad w_{,\tau} = 
{\undertilde u} \, u_{,\tau}\,, \qquad
{\wt v} + {\undertilde w} - {\wt u}\, {\undertilde u} + \frac{n}{u_{,\tau}}=0\,.
\label{eq:DCA}
\end{equation} 
The variables $v$ and $w$ can be easily eliminated from equations 
(\ref{eq:DCA}), leading to the following scalar D$\Delta$E for the variable $u$
\begin{equation}
\left(\frac{n}{u_{,\tau}}\right)_{\!\!,\tau}=
({\wt u}-\underaccent{\wtilde}{\underaccent{\wtilde}{u}})\, 
{\underaccent{\wtilde}{u}}_{,\tau} +
({\underaccent{\wtilde}{u}}-\,{\wt {\wt{u}}}\,) \,{\wt u}_{,\tau}\,. 
\label{eq:scalarDCA}
\end{equation}  
Since by construction the system is symmetric with respect to the lattice 
variables and the associated parameters, the replacement $({\wt \cdot},n,\tau)$ 
with $({\wh \cdot},m,\sigma)$ in equations 
(\ref{eq:compA})-(\ref{eq:scalarDCA}), delivers the matrix $B$ and the 
associated D$\Delta$E for the variables $(u,v,w)$. Explicitly, the matrix $B$ 
takes the dyadic expressions   
\begin{equation}
B= b_i\, b'_j = d_i\,d'_j\,,   \label{eq:dyadicB}
\end{equation}
where
\begin{subequations}\label{eq:compB}
\begin{align}
{b}_i&=(1,{\wh u},{\wh v})\,,  
&{b'}_i&=(m+{\wh v} u_{,\sigma} - {\wh u} w_{,\sigma}\,,\,w_{,\sigma}\,,\,- 
u_{,\sigma})\,,\label{eq:compB1}\\
{d}_i& = (-u_{,\sigma}\,,\,-v_{,\sigma}\,,\,m+\underaccent{\what}{w} u_{,\sigma}
- \underaccent{\what}{u} v_{,\sigma})\,,
&{d'}_i&=(\underaccent{\what}{w},-\underaccent{\what}{u},1)\,, \label{eq:compB2}
\end{align} 
\end{subequations}
$i=0,1,2$, and ${\underaccent{\what}{\cdot}}$ denotes backward shift in the 
lattice variable $m$. The system of the compatible D$\Delta$Es is
\begin{equation}
v_{,\sigma}={\wh u}\, u_{,\sigma},\qquad w_{,\sigma} = 
\underaccent{\what}{u} \, u_{,\sigma}\,,\qquad {\wh v} + \underaccent{\what}{w} 
- {\wh u}\, \underaccent{\what}{u} + \frac{m}{u_{,\sigma}}=0\,, \label{eq:DCB}
\end{equation} 
which by eliminating the variables $v$ and $w$, leads to the following scalar
D$\Delta$E for $u$
\begin{equation}
\left(\frac{m}{u_{,\sigma}}\right)_{\!\!,\sigma}=
({\wh u}-\underaccent{\what}{\underaccent{\what}{u}})\, 
{\underaccent{\what}{u}}_{,\sigma} +
({\underaccent{\what}{u}}-\,{\wh {\wh{u}}}\,) \,{\wh u}_{,\sigma} \,.
\end{equation}
The D$\Delta$Es in equations (\ref{eq:DCA}) and (\ref{eq:DCB}), represent the 
{\em hetero}-B\"ackund transformations between copies of the compatible set of 
PDEs, which we derive in the next section, for different values of the 
parameters $n$ and $m$. This issue, along with the derivation of 
{\em auto}-B\"acklund transformations of the continuous systems \cite{preTN}, 
will be the subject of a future publication.

\section{Compatible partial differential equations}

We now focus on the dependence of $\Psi$, and the dependent variables, on the 
continuous lattice parameters $(\tau,\sigma)\in \mathbb{C}^2$. Since we drop  
for the moment the discrete operations, the variables which are assigned to 
different lattice vertices are interpreted as new dependent variables. 
However, we point out that, we may retain the different lattice labels on the 
associated fields, since by construction the whole set the P$\Delta$E's, 
D$\Delta$E's and PDE's are mutually compatible to each other.

\subsection{The generating PDE of the Boussinesq hierarchy}

The compatibility condition $\Psi_{,\tau \sigma}=\Psi_{,\sigma \tau}$ on the 
linear system (\ref{eq:contLax}), for every value of $\zeta$, leads to the 
following system of matrix PDEs
\begin{equation}
(\sigma-\tau)A_{,\sigma} + [A\,,\,B]=0\,, \qquad (\sigma-\tau) B_{,\tau} + 
[A\,,\,B]=0\,.  \label{eq:compAB}
\end{equation}
Let us now choose the dyadic forms of $A$ and $B$, given by (\ref{eq:compA1}) 
and (\ref{eq:compB1}) respectively, and which for convenience we rewrite below,
along with the relevant trace relations given by equations (\ref{eq:traceAB}),  
\begin{subequations} \label{eq:dyadicAB}
\begin{eqnarray}
A_{ij}=a_i a'_j \,,&\qquad& a^i\,a'_i=n\,, \label{eq:dyadicAB1} \\
B_{ij}=b_i b'_j \,,&\qquad& b^i\,b'_i=m\,, \label{eq:dyadicAB2}
\end{eqnarray}
\end{subequations}
$i,j=0,1,2$. Indices are raised and lowered using the Kronecker symbol 
$\delta_{ij}$, and summation over repeated upper and lower indices is 
understood. From equations (\ref{eq:compAB}), it follows that we may consider a 
more general case where $n$ and $m$ are parameter functions of $\tau$ and 
$\sigma$, respectively. In the following, we adopt the latter option, unless 
otherwise stated. We shall only require that when the relevant discrete 
operations are performed, $n$ and $m$, vary by unit steps only.  

Inserting $A$ and $B$, given by (\ref{eq:dyadicAB}) into equations 
(\ref{eq:compAB}), and after simple algebraic manipulations, taking into
account the fact that $a_0=b_0=1$, we arrive at the following system of PDE's
\begin{subequations}\label{eq:prepdes}
\begin{align}
(\tau-\sigma)\,a_{i,\sigma} - a^j\, b'_j \, (a_i-b_i) &=0,\quad 
&(\tau-\sigma)\,a'_{i,\sigma}& + a^j\, b'_j\, a'_i - b^j a'_j \, b'_i=0\,, 
\label{eq:prepdes1} \\
(\tau-\sigma)\,b_{i,\tau} - b^j\, a'_j\,(a_i-b_i) &=0,\quad
&(\tau-\sigma)\,b'_{i,\tau}& + a^j\,b'_j\, a'_i - b^j a'_j\, b'_i=0\,, 
\label{eq:prepdes2}
\end{align}
\end{subequations}
$i,j=0,1,2$. From the right set of the above of equations we find that 
$a'_{i,\sigma}=b'_{i,\tau}$. Using this fact, and the explicit form of $a'_i$ 
and $b'_i$, given by (\ref{eq:compA1}) and (\ref{eq:compB1}) respectively, we 
introduce the variables $(f_0,f_1,f_2)$, by the relations
\begin{equation}
f_1=w\,,\quad f_2=-u\,,\quad 
f_{0,\tau}= n - a^i\,f_{i,\tau}\,,\quad f_{0,\sigma}= m - b^i\,f_{i,\sigma} \,,
\label{eq:deffis}
\end{equation}
$i=1,2$, and which satisfy
\begin{equation}
a'_i = f_{i,\tau}\,,\qquad b'_i = f_{i,\sigma}\,, \label{eq:potenf}
\end{equation} 
$i=0,1,2$. In terms of the new variables $f_i$, the system of equations 
(\ref{eq:prepdes}) reads
\begin{subequations}\label{eq:fieldeqs}
\begin{eqnarray} 
(\tau-\sigma) a_{i,\sigma} - \big(m + (a^j-b^j)\,\, f_{j,\sigma}\big) (a_i-b_i) 
&=&0\,, \label{eq:fieldeqs1} \\
(\tau-\sigma) b_{i,\tau} - \big(n - (a^j-b^j)\,\, f_{j,\tau}\big) (a_i-b_i)&=&0
\,, \label{eq:fieldeqs2} \\
(\tau-\sigma) f_{i,\tau\sigma} + m f_{i,\tau} - n f_{i,\sigma} + C_{ij}(a^j-b^j)
&=&0\,, \label{eq:fieldeqs3}   
\end{eqnarray}
\end{subequations}
where $i,j=1,2$. The components of the matrix $(C_{ij})$, which is in general 
a rank two matrix, are defined as follows 
\begin{equation}
C_{ij}=f_{i,u}\,f_{j,v} + f_{i,v}\,f_{j,u}\,, \label{eq:defC}
\end{equation}
$i,j=1,2$. Solving the linear system (\ref{eq:fieldeqs3}) for
the differences $(a_i-b_i)$, and using the remaining field equations 
(\ref{eq:fieldeqs1}), (\ref{eq:fieldeqs2}), we obtain a fourth order coupled 
system of PDE's for the variables $(f_1,f_2)$, which is omitted because of its
length. The latter system of PDEs are the Euler-Lagrange equations for the 
variational problem, associated with the Lagrangian density
\begin{equation} 
L = (\tau-\sigma)\,\frac{F E}{J^2} + m(\sigma)\, \frac{F}{J} + n(\tau)\, 
\frac{E}{J}\,. \label{eq:lagrangian}
\end{equation}
Here, the scalar quantities $F,E$ and $J$ are given by
\begin{equation} 
F= {{\boldsymbol f}}_{\!\!,\tau\sigma} \times {{\boldsymbol f}}_{\!\!,\tau}\,,
\quad      
E= {{\boldsymbol f}}_{\!\!,\tau\sigma} \times {{\boldsymbol f}}_{\!\!,\sigma}\,,
\quad
J= {{\boldsymbol f}}_{\!\!,\tau} \times {{\boldsymbol f}}_{\!\!,\sigma}\,, 
\label{eq:FJE}
\end{equation} 
where ${\boldsymbol f}$ is the vector with components $(f_1,f_2)$. In the 
following, we denote the parameter family of fourth order PDEs for the variables
$(f_1,f_2)$, by ${\boldsymbol E}(L)=0$. We conclude this section by noticing 
that the Lagrangian $L$ remains invariant, modulo $L$ and null Lagrangians, 
under the action of the projective group of linear fractional transformations on
$(f_1,f_2)$, i.e.
\begin{equation} 
{(f_1,f_2)}\, \mapsto\, 
\left(\frac{{\mathcal A}_{11} f_1 + {\mathcal A}_{12} f_2 +{\mathcal A}_{13}}
{{\mathcal A}_{31} f_1 + {\mathcal A}_{32} 
f_2 +{\mathcal A}_{33}}\,,\, 
\frac{{\mathcal A}_{21} f_1 + {\mathcal A}_{22} f_2 +{\mathcal A}_{23}}
{{\mathcal A}_{31} f_1 + {\mathcal A}_{32} 
f_2 +{\mathcal A}_{33}}\right)\,, \label{eq:symm1}
\end{equation}
where $({\mathcal A}_{ij})\in\mathrm{SL}(3,\mathbb{C})$. If in addition the 
parameter functions $n$ and $m$ are constants, then $L$ remains invariant under 
the affine base transformations 
\begin{equation} 
(\tau,\sigma) \mapsto (\varepsilon_1 \tau + \varepsilon_2,\,\,\varepsilon_1 
\sigma + \varepsilon_2)\,, 
\,\qquad \varepsilon_1,\varepsilon_2 \in 
\mathbb{C},\,\, \varepsilon_1 \neq 0\,. \label{eq:symm2}
\end{equation}
The symmetry transformations of $L$, given by (\ref{eq:symm1}) and 
(\ref{eq:symm2}), fall into the class of the so-called divergence symmetries of 
the associated variational problem. Every symmetry of this type is inherited as 
a Lie-point symmetry, to the corresponding Euler-Lagrange equations, see 
\cite{Olver1} for a detailed account on these topics.
  
\subsection{The Boussinesq hierarchy}

The coupled system of PDEs ${\boldsymbol E}(L)=0$, encodes the complete 
hierarchy of the Boussinesq equations. This can be illustrated, by imposing on 
$\Psi$ and on the associated dependent variables, an additional dependence on an
infinite set of indeterminates $x^i=(x^1,x^2,\ldots)$. Following the 
construction given in \cite{Nij2}, \cite{Nij3}, the dependence is given 
explicitly by certain (infinite) differential relations, in such a way that all 
these relations are mutually consistent to each other. To this end, we introduce
the matrix $H$ with components
\begin{equation}
H_{ij,\tau} = a_i\,f_{j,\tau}\,,\quad H_{ij,\sigma} = a_i \,f_{j,\sigma}\,, 
\label{eq:defH}
\end{equation}
$i,j=0,1,2$, and the matrix $Z$ by the relation
\begin{equation}
{Z} = \left( 
\begin{array}{ccc}0&1&0\\ 0&0&1 \\ -\zeta &0&0 \end{array} \right)\,. 
\label{eq:defZ}
\end{equation}
The existence of the matrix $H$ for the variables $(\tau,\sigma)$, may be easily
verified by using the system of PDE's (\ref{eq:fieldeqs}). In terms of the above
matrices, the linear differential relations for $\Psi$ which can be 
simultaneously imposed on the linear system (\ref{eq:contLax}), are given by
\begin{subequations}\label{eq:BSQhier}
\begin{eqnarray}
\partial_{1} \Psi &=& ({Z}\,\,\, +\,[\,\,\partial_{\zeta}{Z}\,, {H}\,]\,)\Psi\,, 
\label{eq:BSQhier1}\\  
\partial_{2} \Psi &=& ({Z}^2 +\, [\,{\partial_{\zeta}{Z}^2}, {H}\,]\,)\Psi\,, 
\label{eq:BSQhier2} \\
\partial_{i+3} \Psi &=& -(\zeta \partial_{i}\Psi\, + \,\partial_{i}{H})\,\Psi\,, 
\label{eq:BSQhier3}
\end{eqnarray}
\end{subequations}   
where $\partial_i=\partial_{x^i}$ and $i\in \mathbb{Z}/3 \mathbb{Z}$.

Taking the compatibility condition 
$\partial_1\partial_2\Psi=\partial_2\partial_1\Psi$, on equations 
(\ref{eq:BSQhier1}) and (\ref{eq:BSQhier2}), we obtain a set of PDE's for the 
various entries of the matrix $H$. A straightforward differentiation and 
elimination process applied to the latter system, and identifying the main field
variables $(f_1,f_2)$ using (\ref{eq:defH}), we end up with the following set of
PDE's
\begin{subequations}\label{eq:preBSQ}
\begin{eqnarray} 
\partial_{1}\partial_{2}f_1 &=& \textstyle{2 (\partial_2)^2 f_2 - 
3 \partial_1 (\partial_1 f_2)^2
- \frac{1}{2} (\partial_1)^2 \partial_{2} f_2 
+ \frac{1}{2} \partial_1 \partial_2(f_2^2) + \frac{1}{4} (\partial_1)^4 f_2 }\,,
\label{eq:preBSQ1} \\ 
\partial_{1}\partial_{1} f_1 &=& 
\textstyle{ \frac{1}{2}\partial_{1}\partial_{2} f_2 + \frac{1}{2} \partial_{1} 
(\partial_{1} f_2)^2  - \frac{1}{2}(\partial_{1})^3 f_2}\,.  \label{eq:preBSQ2}
\end{eqnarray}
\end{subequations}
Then, the final compatibility condition  
$(\partial_1)^2 \partial_2 f=\partial_2 (\partial_1)^2 f$ on the above system, 
leads to the Boussinesq equation
\begin{equation}
3 (\partial_2)^2 {\mathcal Q}+ (\partial_1)^4 {\mathcal Q} + 6 (\partial_1)^2 
({\mathcal Q}^2) = 0 \, , \label{eq:BSQ}
\end{equation}
where ${\mathcal Q}=-\partial_1 f_2$. Similarly, higher members of the 
Boussinesq hierarchy can be derived from the system (\ref{eq:BSQhier}), in a 
systematic way.

\subsection{The Ernst equations for an Einstein-Maxwell-Weyl field}

Now we discuss various subsystems which are included into the equations 
${\boldsymbol E}(L)=0$. Such subsystems arise, by imposing additional algebraic 
or differential constraints on the space of the independent and the dependent 
variables and on the parameter space, in a consistent way with the equations 
${\boldsymbol E}(L)=0$, or equivalently, with the auxiliary system 
(\ref{eq:fieldeqs}).  

Assuming first that the independent variables are real, one such consistent 
algebraic constraint is given by the following set of relations on the auxiliary
variables,  
\begin{equation}
a_1=b_1 + \frac{(\sigma-\tau) f_1^\star}{f_2 + f_2^\star+ f_1\,f_1^\star}\,,
\qquad
a_2=b_2 + \frac{\sigma-\tau}{f_2 + f_2^\star+ f_1\,f_1^\star}\,, 
\label{eq:constr1}
\end{equation}
and on the parameters
\begin{equation}
n + n^\star + 1=0\,,\quad m + m^\star + 1=0\,, \label{eq:constr2}
\end{equation}
where $\star$ denotes complex conjugation. Imposing the constraints 
(\ref{eq:constr1}) and (\ref{eq:constr2}) on  the auxiliary system 
(\ref{eq:fieldeqs}), the latter reduces to a second order coupled system of PDEs
for the main dependent variables $(f_1,f_2)$ solely, which is given by equations
(\ref{eq:fieldeqs3}). Relabeling the variables $(f_1,f_2)$ by 
$(\varPhi,\mathcal{E})$, and using the language of forms, the latter set of PDEs
can be written as
\begin{subequations}\label{eq:Ernst}
\begin{eqnarray}
({\mathcal E}+{\mathcal E}^\star + \varPhi \varPhi^\star) 
\big(\dif (\varrho \ast \dif {\mathcal E}) - {\rm i}\, \boldsymbol{\beta} 
\wedge \dif {\mathcal E}\big)= 
2 \varrho (\dif {\mathcal E} + \varPhi^\star \dif \varPhi)\wedge \ast 
\dif {\mathcal E},
\label{eq:Ernst1}\\
({\mathcal E}+{\mathcal E}^\star + \varPhi \varPhi^\star) 
\big(\dif (\varrho \ast \dif \varPhi) - {\rm i}\, \boldsymbol{\beta} \wedge 
\dif \varPhi\big)= 2 \varrho (\dif {\mathcal E} + \varPhi^\star \dif \varPhi)
\wedge \ast \dif \varPhi\, .\label{eq:Ernst2}
\end{eqnarray}
\end{subequations}
Here, $\ast$ denotes the Hodge duality operator acting on the basis of one 
forms by
\begin{equation}
\ast \dif \tau = \dif \tau,\quad \ast \dif \sigma = -\dif \sigma\,,
\end{equation} 
while the function $\varrho$, and the one-form $\boldsymbol{\beta}$ are defined 
as 
\begin{equation}
\varrho=\frac{1}{2}(\sigma-\tau)\,,\qquad \boldsymbol{\beta}=\nu(\tau)\dif \tau 
+ \mu(\sigma) \dif \sigma \,,
\end{equation}
where $\nu(\tau)$ and $\mu(\sigma)$, are real parameter functions of the 
indicated arguments. The system of PDEs (\ref{eq:Ernst}) forms the basis of the 
Einstein's field equations for plane symmetric spacetimes, in the presence of a 
source-free Maxwell field and a Weyl neutrino field, and they are known in 
general relativity as the Ernst equations for an Einstein-Maxwell-Weyl field, 
\cite{Ernst1}-\cite{Alex1}. 

\section{Symmetry reduction to a second order system of ODEs}

We are now interested in finding particular similarity solutions of the 
equations $\boldsymbol{E}(L)=0$. This will be achieved by, first, fixing the 
parameters functions $n$, $m$ to be constants, and then by choosing a specific 
one-dimensional subgroup from the full group of Lie-point symmetries 
(\ref{eq:symm1}), (\ref{eq:symm2}). 
The requirement of invariance of the solutions under the one-dimensional 
subgroup, leads to the corresponding similarity solutions.
 
Our motivation in finding such solutions is two-fold. First, one hopes to 
discover some explicit solutions of the fourth order coupled system 
$\boldsymbol{E}(L)=0$, which would provide physically significant solutions, 
in connection with the Ernst equations. In the present work however, we are 
interested mostly in writing down the system of the reduced ordinary 
differential equations (ODE), since the latter potentially define higher order 
analogous of the Painlev\'e ODE's. 

To be more specific, we shall be interested in finding solutions of the 
equations $\boldsymbol{E}(L)=0$, which remain invariant along the orbits 
generated by the following vector field 
\footnote{The symmetry generator (\ref{eq:symmgen1}), generalizes the one used 
in \cite{NHJ}, \cite{TTX}, for the symmetry reduction of a scalar subsystem, 
included into equations $\boldsymbol{E}(L)=0$, to the full Painlev\'e VI 
equation.}   
\begin{equation}
X = \tau\partial_\tau + \sigma\partial_\sigma + 2\alpha_1 f_1 \partial_{f_1} 
+ 2\alpha_2 f_2 \partial_{f_2}\,,
\label{eq:symmgen1} 
\end{equation}
where $\alpha_1,\,\alpha_2$, arbitrary complex parameters. Requiring invariance,
means that the function $(f_1,f_2)$ should satisfy the equations 
$\boldsymbol{E}(L)=0$, and also the compatible differential constraints 
\begin{equation}
\tau f_{i,\tau} + \sigma f_{i,\sigma} = 2 \alpha_i f_i \qquad i=1,2\,, 
\label{eq:diffcon} 
\end{equation}
The general solution of the latter PDEs is
\begin{equation}
f_i(\tau,\sigma) = F_i\left(t\right) (\tau \sigma)^{\alpha_i}\,, \qquad 
t= \frac{\tau}{\sigma}\,, \qquad i=1,2\,. 
\qquad {\mbox{(no summation)}} \label{eq:invf}
\end{equation}
Inserting the main variables in the form given by equation (\ref{eq:invf}), 
into equations $\boldsymbol{E}(L)=0$, we obtain a fourth order system of ODEs 
for the functions $F_i$, which is omitted because of its length. However, an 
effective way in dealing with the latter system of ODEs, is to consider the 
auxiliary system (\ref{eq:fieldeqs}). To this end, one needs first to find the 
vertical components of the vector field $X$, which are associated with the 
auxiliary variables $a_i$, $b_i$, and subsequently the consistent invariant 
form of the variables $a_i$, $b_i$. The corresponding symmetry generator 
${\mathcal X}$ for the equations (\ref{eq:fieldeqs}), is  
\begin{equation}
{\mathcal X} = X + (1-2 \alpha_1) (a_1 \partial_{a_1} + b_1 \partial_{b_1}) + 
(1-2 \alpha_2)(a_2 \partial_{a_2} + b_2 \partial_{b_2})\,, \label{eq:symmgen2} 
\end{equation}  
which leads us to conclude that the invariant form of the auxiliary potentials 
is
\begin{equation} \begin{array}{l}
a_i(\tau,\sigma) = P_i\left(t\right) (\tau \sigma)^{1/2 - \alpha_i}\,,\\ 
b_i(\tau,\sigma) = R_i\left(t\right) (\tau \sigma)^{1/2 -\alpha_i}\,, 
\end{array}
\qquad t= \frac{\tau}{\sigma}\,, \qquad i=1,2\,.  \qquad {\mbox{(no summation)}}
\label{eq:invab}
\end{equation}   
Substituting the dependent variables $a_i$, $b_i$ and $f_i$ given by 
(\ref{eq:invab}) and (\ref{eq:invf}), respectively, into equations 
(\ref{eq:fieldeqs}), we obtain the following system of ODEs
\begin{subequations} \label{eq:odes}
\begin{align}
-(t-1) \left(t P'_i - (\textstyle{\frac{1}{2}}-\alpha_i) P_i\right) &= 
\left(m - t^{1/2} (P^j-R^j) (t F'_j - \alpha_j F_j ) \right)(P_i-R_i)\,, 
\label{eq:odes1} \\
(t-1) \left(t R'_i + (\textstyle{\frac{1}{2}}-\alpha_i) R_i\right) &= 
\left(n t - t^{1/2} (P^j-R^j) (t F'_j + \alpha_j F_j ) \right)(P_i-R_i)\,, 
\label{eq:odes2} \\
t^2 (1-t) F''_i + t\big( (n-1)t+m+1\big)F'_i &- \alpha_i \big( (n-\alpha_i) t-
(m-\alpha_i)\big) F_i = \nonumber \\ 
& 2\, t^{1/2} \left(t^2 F'_i F'_j - \alpha_i F_i \alpha_j F_j\right) 
\left(P^j-R^j\right)\,, \label{eq:odes3}
\end{align}
\end{subequations}
where prime denotes differentiation with respect to $t$. Employing a general 
quadratic {\em Ansatz} on the first order derivatives of $F_i$'s, we find that 
the system (\ref{eq:odes}), admits two first integrals. Denoting the complex 
constants of integration by $s_1$ and $s_2$, these first integrals are given by  
\begin{align} \label{eq:int1}
&(P_1-R_1)^2\big(t^2 {F'_1}^2 - {\alpha_1}^2 {F_1}^2\big) + 
(P_1-R_1)(P_2-R_2)(t^2 F_1' F_2' - {\alpha_1} {\alpha_2} F_1 F_2)& \nonumber \\ 
&+t^{1/2}\left[ \big(F_1' +\frac{{\alpha_1}}{t} F_1\big) \big((2 {\alpha_1} -1) 
P_1 -m(P_1-R_1)\big) - t \big(F_1' -\frac{{\alpha_1}}{t} F_1\big) 
\big((2{\alpha_1} -1) R_1 +n (P_1-R_1)\big) \right]& \nonumber \\
&+t ({\alpha_1} F_1 F_2'-{\alpha_2} F_1' F_2)\left[ (P_1 R_2-P_2 R_1) 
\frac{{\alpha_1}+{\alpha_2}-1}{{\alpha_1}-{\alpha_2}}+
(R_1 R_2 - P_1 P_2) \right] = s_1\,, &
\end{align}
together with the same relation, obtaind from (\ref{eq:int1}) by interchanging 
the subscripts $1\leftrightarrow 2$. Through a straightforward elimination and 
differentiation process applied to the system (\ref{eq:odes}), taking into 
account the first integrals and introducing new dependent variables 
$({\mathcal G}_1,{\mathcal G}_2)$ by the substitutions, 
\begin{equation}
{ {\mathcal G}_i(t)} = \frac{t}{\alpha_i}\,\frac{F'_i(t)}{F_i(t)} \,, 
\qquad i=1,2\,,  \label{eq:sub} 
\end{equation}  
we end up with a second order, coupled system of ODEs for the variables 
$({\mathcal G}_1,{\mathcal G}_2)$, which involves six free parameters, namely 
$(n,m,\alpha_1,\alpha_2,s_1,s_2)$. In order to give an explicit form of the 
latter system of ODEs, we introduce the auxiliary quantities $(Q_1,Q_2)$ as 
follows
\begin{eqnarray} \label{eq:Q1}
Q_1= 
\frac{m + n t}{2 \alpha_1 ({\mathcal G}_1 - {\mathcal G}_2)} +
\frac{(m - n t) {\mathcal G}_2}{2 \alpha_1 ({\mathcal G}_1 - {\mathcal G}_2)} +
\frac{(t - 1) \alpha_2 {\mathcal G}_1 {\mathcal G}_2 (1 - {{\mathcal G}_2}^2)}
{2 \alpha_1 ({\mathcal G}_1 - {\mathcal G}_2)^2} - 
\frac{(t - 1)(\alpha_1 - \alpha_2 ) ({{\mathcal G}_2}^2 - 1)}
{2 \alpha_1 ({\mathcal G}_1 - {\mathcal G}_2)^2} + \nonumber \\
\frac{(t - 1) {{\mathcal G}_1}^2 ({{\mathcal G}_2}^2 - 1)} {2 ({\mathcal G}_1 - 
{\mathcal G}_2)^2} + \frac{t (t - 1) ({{\mathcal G}_2}^2 - 1) {\mathcal G}'_1} 
{2 \alpha_1 ({\mathcal G}_1 - {\mathcal G}_2)^2} - \frac{t (t - 1) 
({\mathcal G}_1 {\mathcal G}_2 - 1) {\mathcal G}_2'}{2 \alpha_1 ({\mathcal G}_1- 
{\mathcal G}_2)^2}\,,
\end{eqnarray}
and $Q_2$ is obtained from $Q_1$ by interchanging the subscripts 
$1\leftrightarrow 2$.  Then, the aforementioned system of ODEs is given by
\begin{align}\label{eq:ode1} 
&\frac{\alpha_1 \,\alpha_2}{\alpha_1-\alpha_2} ({\mathcal G}_1 - 
{\mathcal G}_2) (Q_1 Q'_2 - Q'_1 Q_2) + \big((t - 1){\mathcal G}_1 - 
(t+1)\big) {Q'_1} \nonumber & \\ & -\frac{{\alpha_1}^2}{t}\left( 
({\mathcal G}_1-1)^2 - \frac{4{\mathcal G}_1}{t - 1}\right){Q_1}^2 
-\frac{\alpha_1\, \alpha_2}{t}\left( ({\mathcal G}_2-1)^2 - 
\frac{4 {\mathcal G}_2}{t - 1} - \frac{\alpha_1 ({\mathcal G}_1-
{\mathcal G}_2)^2}{\alpha_1-\alpha_2} \right) Q_1 Q_2 & \nonumber \\ & 
-\frac{{\alpha_1}^2}{t}\left((t-1)({\mathcal G}_1-1)^2 - 4 {\mathcal G}_1 + 
\frac{t}{\alpha_1}({\mathcal G}_1-1) +\frac{2 t}{\alpha_1} \frac{(n+m)}{(t - 1)}
\right)  Q_1 = s_1\,,  &
\end{align}
together with the same equation obtained from equation (\ref{eq:ode1}), by 
making the usual interchanges in the subscripts.

\section{Perspectives}

We finish our discussion by noting that, on the continuous level, the generating
PDE of the Boussinesq hierarchy, namely the equations $\boldsymbol{E}(L)=0$, 
include the one of the KdV hierarchy. This is achieved by considering the case, 
where the dependent variables $(f_1,f_2)$ coincide, i.e. $f:=f_1=f_2$. The 
generating PDE of the KdV hierarchy was given in \cite{NHJ}, and is the 
Euler-Lagrange equation for the variational problem associated with the
Lagrangian
\begin{equation} 
{\mathcal L} = (\tau-\sigma) \frac{(f_{,\tau\sigma})^{\,2}}{f_{,\tau}f_{,\sigma}} + 
\frac{1}{\tau-\sigma} \left( m^2 \frac{f_{,\tau}}{f_{,\sigma}} + 
n^2 \frac{f_{,\sigma}}{f_{,\tau}} \right)\, . \label{eq:KdV}
\end{equation}
and which we denote by $\boldsymbol{E}({\mathcal L})=0$.
The Lagrangian ${\mathcal L}$ can be reduced from the Lagrangian $L$ given by 
equation (\ref{eq:lagrangian}), considering the aforementioned case, in a 
consistent way with the reduced equations. In \cite{NHJ}, cf \cite{TTX}, 
it was shown that a specific symmetry reduction of the equations 
$\boldsymbol{E}({\mathcal L})=0$ leads to invariant solutions, built from the 
full Painlev\'e VI equation. Thus, we expect that there exists an intimate 
connection of the six-parameter system of coupled ODEs, presented in Section 7, 
with the hierarchies of the Painlev\'e VI and the Garnier systems 
\cite{Garnier}. The latter issue along with the isomonodromy problem of the 
relevant ODEs, will be left to a future study.    

The derivation of generating PDEs, which, in turn, provide a mechanism for 
obtaining new integrable systems, is a project of equal importance. To this end,
the classification of scalar discrete integrable systems, obtained recently in 
\cite{Adler}, could serve as a starting point. However, some of the extra 
restrictions imposed on the classification scheme in \cite{Adler}, would 
exclude, in a strict sense, the discrete BSQ system (\ref{eq:latticeBSQ}). 
Thus, the classification of quadrilateral integrable discrete systems, in more 
than one dependent variable, seems to be a more complicated problem. 

In the light of the intimate connection between the PDE's 
$\boldsymbol{E}({L})=0$ and $\boldsymbol{E}({\mathcal L})=0$, and the Ernst 
equations, the former fourth order integrable systems acquire a certain physical
importance. This stems from the fact that the solution space of the hyperbolic 
Ernst equations is embedded into the solution space of the fourth order systems. 
Consequently, it would be interesting to consider to what extent, solutions of 
the larger continuous systems and their discrete compatible systems, could 
provide new exact solutions of the Einstein's equations, for various physically 
significant problems. Such a problem is the head-on collision of plane 
fronted gravitational waves, coupled with electromagnetic and neutrino waves, 
see \cite{Griffithsbook} and references therein.

\subsection*{Acknowledgements} 
The work of A.T. was supported by the
European postdoctoral fellowship Marie Curie, contract No HPMF-CT-2002-01639.

\end{document}